\documentclass[aps, prl, superscriptaddress, twocolumn, amsfonts, amsmath, amssymb, floatfix]{revtex4-1}

\usepackage{graphicx}
\usepackage{dcolumn}
\usepackage{bm}
\usepackage{epsfig}
\usepackage{braket}
\usepackage{color}
\usepackage{ulem}
\usepackage{amsmath}
\usepackage[colorlinks=true, letterpaper=true, pdfstartview=FitV, linkcolor=blue, citecolor=blue, urlcolor=blue]{hyperref}

\begin{document}

\title{Quantum fluctuations  and  Gross-Pitaevskii theory}

\author{Sandro Stringari}
\affiliation{INO-CNR BEC Center and Dipartimento di Fisica, Universit\`a di Trento, 38123 Povo, Italy}

\date{\today}

\begin{abstract}
Using the linearized version of the time dependent Gross-Pitaevskii equation  we calculate the dynamic response of a Bose-Einstein condensed gas  to periodic density and particle perturbations.  The zero temperature limit of the fluctuation-dissipation theorem is    used to evaluate the corresponding quantum fluctuations induced by the elementary excitations in the ground state. In uniform conditions the predictions of Bogoliubov theory, including the infrared divergency of the particle distribution function and the quantum depletion of the condensate, are exactly reproduced by Gross-Pitaevskii theory. Results are also given for the crossed particle-density response function  and the extension of the formalism to  non uniform systems is discussed.  The generalization of the Gross-Pitaevskii equation to include beyond mean field effects is finally considered and an explicit result for the chemical potential is found, in agreement with the prediction of Lee-Huang-Yang theory.

\end{abstract}
	
\maketitle

\textit{\bf Introduction.}

Bogoliubov \cite{Bog} and Gross-Pitaevskii \cite{G,P} theories represent basic approaches to the physics of a weakly interacting Bose gas. While Bogoliubov theory is based on a  quantum description   where the particle operators are transformed into quasi-particle operators, allowing for an explicit diagonalization of the quantum Hamiltonian, Gross-Pitaevskii theory consists of an equation for  the order parameter, a classical field associated with the spontaneous breaking of gauge symmetry.  

The main purpose of the present paper is to show that the quantum fluctuations exhibited by an interacting Bose-Einstein condensate, can be properly calculated using the formalism of time dependent Gross-Pitaevskii theory (TDGP), recovering   the results of Bogoliubov theory and allowing for applications to non uniform configurations.  In addition to the density fluctuations an important case considered in this work concerns the particle fluctuations whose knowledge gives access to  the momentum distribution and to the quantum depletion of the condensate.
 We will also develop a  generalization of the Gross-Pitaevskii equation for the order parameter, accounting for beyond mean field effects.

We will make explicit use of the fluctuation dissipation theorem \cite{Kubo}, which relates the fluctuations associated with a given physical operator $\hat F$ to the imaginary part of the corresponding dynamic polarizability.
At zero temperature the theorem takes the form (see, for example, \cite{book})
\begin{equation}
\langle \{(F^\dagger-\langle F^\dagger\rangle), (F-\langle F \rangle)\}  \rangle =\frac{\hbar}{\pi}\int_{-\infty}^{+\infty}d\omega \chi_F^{''}(\omega)sign(\omega)
\label{FD}
\end{equation}
where $\{\hat{A},\hat{B}\} \equiv \hat{A}\hat{B}+\hat{B}\hat{A}$ is the anticommutator between the two operators. Identity (\ref{FD}) emphasizes the quantum nature of the fluctuations \cite{FD}. Equivalently, one can also write (again at zero temperature)
\begin{equation}
\langle (F^\dagger-\langle F^\dagger\rangle) (F-\langle F \rangle)  \rangle =\frac{\hbar}{\pi}\int_0^{+\infty}d\omega \chi_F^{''}(\omega)
\label{FD+}
\end{equation} 

The crucial ingredient entering Eqs. (\ref{FD},\ref{FD+}) is the dynamic polarizability, defined 
by the variation 
\begin{equation}
\delta \langle {\hat F}^\dagger \rangle=\lambda e^{\eta t} [ e^{-i\omega t}  \chi_F(\omega) + e^{i\omega t} \chi_{F^\dagger}(-\omega)]
\label{chi}
\end{equation}
of the average value of the operator $F^{\dagger}$, induced by an external time dependent perturbation of the form 
\begin{equation}
H_{pert} = -\lambda e^{\eta t} \left({\hat F} e^{-i\omega t}+{\hat F}^\dagger e^{+i\omega t}\right)
\label{pert}
\end{equation}
with $\eta$ positive and small, ensuring that at $t=-\infty$ the system is governed by the unperturbed Hamiltonian.  Perturbation theory yields the following result for the dynamic polarizability at zero temperature \cite{Kubo}:
\begin{align}
&\chi_F(\omega) \equiv \chi_{{\hat F}^\dagger, {\hat F}} =\noindent \\
&-\frac{1}{\hbar} \sum_{n}\left[\frac{\langle 0|{\hat F}^\dagger|n\rangle \langle n|{\hat F}|0\rangle}{\omega-\omega_{n0}+i\eta}-\frac{\langle 0|{\hat F}|n\rangle\langle n |{\hat F}^\dagger|0\rangle}{\omega+\omega_{n0}+i\eta}\right] \; .
\label{kubo2}
\end{align} 
If the operator $\hat F$ does not conserve the total number of particles it is convenient to use the grand canonical formalism, adding the term $-\mu \hat{N}$  to the unperturbed  Hamiltonian. 
 
The time dependent Gross-Pitaevskii theory is well suited to calculate the  response function $\chi(\omega)$ and consequently  provides direct access to the  quantum fluctuations of the operator $\hat F$, through the use of Eqs. (\ref{FD},\ref{FD+}). An important example are the density fluctuations associated with the $\bf q$-component  
${\hat\rho}_{\bf q}= \sum_{\bf p} \hat{a}^\dagger_{{\bf p}-\hbar {\bf q}}\hat{a}_{\bf p}$ 
of the density operator, where  $\hat{a}^\dagger$ and $\hat{a}$ the usual creation and annihilation particle operators. In this case   Eq. (\ref{FD}) gives access to the  density fluctuations and in particular to the  static structure factor
\begin{equation}
S(q)=  \frac{1}{N}\langle \hat{\rho}_{\bf q}^\dagger\hat{\rho}_{\bf q}\rangle - \frac{1}{N}|\langle {\hat\rho_{\bf q}}\rangle|^2 \; .
\label{Sq}
\end{equation} 
 Another important case that will be  discussed in the paper concerns the   fluctuations of the particle operator ${\hat a}_{\bf p}$, where ${\bf p}$ is the momentum of the particle. In this case the left hand side of Eq.(\ref{FD+}) allows for the calculation of  the particle distribution
\begin{equation}
n_p= \langle {\hat a}^\dagger_{\bf p}{\hat a}_{\bf p}  \rangle 
\label{np}
\end{equation}
which, in the presence of Bose-Einstein condensation,  is  known to exhibit an infrared divergent behavior at small momenta \cite{Gavoret}   and whose integral allows for the calculation of  the quantum depletion of the condensate. At first sight it may look surprising that an {\it apparently} classical approach, like Gross-Pitaevskii theory, accounts for these crucial quantum fluctuations. Actually the quantum nature of TDGP  theory   is implicitly taken into account by the use of the fluctuation dissipation theorem.

\textit{\bf Fluctuations in Bogoliubov theory.}

Bogoliubov theory is usually applied to uniform configurations employing the Bogoliubov prescription $\hat{a}_0=\hat{a}^\dagger_0 \equiv \sqrt{N_0}$ where $\hat{a}_0$ and $\hat{a}^\dagger_0$ are the particle annihilation and creation operators relative to the single-particle state ${\bf p}=0$, where Bose-Einstein condensation takes place and $N_0\sim N$ is the number of atoms in the condensate. The Bogoliubov prescription corresponds the assuming the spontaneous breaking of gauge symmetry. It is applied to the grand canonical quantum Hamiltonian
\begin{equation} 
{\hat H} = \int d{\bf r} \left[-{\hat \Psi}^\dagger\frac{\hbar^2 \nabla^2}{2m}{\hat \Psi}+\frac{1}{2} g {\hat \Psi}^\dagger{\hat \Psi}^\dagger{\hat \Psi}\hat{\Psi} -\mu{\hat \Psi}^\dagger{\hat \Psi} \right] \; ,
\label{H}
\end{equation}
after writing the field operator $\hat \Psi$ in terms of the particle annihilation operators:
\begin{equation}
{\hat{\Psi}}= \frac{1}{\sqrt{V}} \sum_{\bf p}e^{i{\bf p} \cdot {\bf r}/\hbar}\hat{a}_{\bf p}
\end{equation}
and keeping only terms quadratic in ${\hat a}_{\bf p}$, ${\hat a}^\dagger_{\bf p}$. The interaction coupling constant $g$ entering the Hamiltonian (\ref{H}) is related to the 3D $s$-wave scattering length by $g=4\pi \hbar^2 a/m$. 

By introducing the Bogoliubov transformations
\begin{align}
&{\hat a}_{\bf p}= u_{\bf p}{\hat b}_{\bf p}+v^*_{-\bf p}{\hat b}^\dagger_{-\bf p} \nonumber \\
&{\hat a}_{\bf p}^\dagger= u_{\bf p}^*{\hat b}_{\bf p}^\dagger+v_{-\bf p}{\hat b}_{-\bf p} \; ,
\label{transf}
\end{align}
which transform particle  ($\hat{a}_{\bf p}$, $\hat{a}^\dagger_{\bf p}$) into quasi-particle ($\hat{b}_{\bf p}$, $\hat{b}^\dagger_{\bf p}$) operators,  
the many-body Hamiltonian (\ref{H}) can   be recast in the diagonal form
\begin{equation}
{\hat H} +\mu N= E_0 + \sum_{\bf p}\epsilon(p){\hat b}^\dagger_{\bf p}{\hat b}_{\bf p} \; ,
\label{ELHY+}
\end{equation}
where 
\begin{equation}
\epsilon(p) = \sqrt{\frac{gn}{m}p^2+\left(\frac{p^2}{2m}\right)^2}
\label{Bogspectrum}
\end{equation}
is the most famous Bogoliubov   spectrum of the elementary excitations  fixed by the interaction coupling constant 
$g$, with $n$ the density of the system, while $E_0$ is the  ground state energy, whose evaluation requires a proper renormalization of the coupling constant in order to avoid the occurrence of   ultraviolet divergencies \cite{LHY}. The excitation spectrum $\epsilon({\bf p})$ exhibits the typical phononic dispersion $\epsilon(p)=cp$ at small momenta, with the sound velocity given by  $c=\sqrt{gn/m}$, and the single particle dispersion $p^2/2m$ at high momenta.    The values of the Bogoliubov amplitudes which diagonalize the Hamiltonian, are given by
\begin{equation}
u_{\bf p}, v_{-\bf p}= \pm \sqrt{ \frac{p^2/2m + gn}{2\epsilon(p)}\pm \frac{1}{2}}
\label{uvB}
\end{equation}
and satisfy the normalization condition $|u_{\bf p}|^2- |v_{-\bf p}|^2=1$. In the Bogoliubov approach 
the elementary excitation carrying momentum ${\bf p}$ is created by the operator ${\hat b}^\dagger_{\bf p}$ applied to the ground state, which is defined as the  vacuum of quasi-particles:
\begin{equation}
\hat{b}_{\bf p}|0\rangle_{Bog} = 0
\end{equation}
for any ${\bf p}\ne 0$.
As a consequence, the density and particle fluctuations in the ground state are straightforwardly calculated by using the Bogoliubov transformations (\ref{transf}) and the commutation rule $[\hat{b}_{\bf  p},\hat{b}^\dagger_{\bf p}]=1$. For example, using the Bogoliubov prescription and approximating $N_0$ with $N$ we can write the density operator in the form ${\hat F}= \rho_{\bf q}= \sqrt{N}({\hat a}_{\bf p}+{\hat a}^\dagger_{-\bf p})$ with ${\bf p}=\hbar {\bf q}$, yielding the result
\begin{equation}
 \langle \hat{\rho}^\dagger_{{\bf q}}{\hat\rho}_{\bf q}\rangle=  N\frac{\hbar^2q^2/2m}{ \epsilon(\hbar q)}
\label{rhorho}
\end{equation}
for the density fluctuations in uniform conditions.  Choosing $\hat{F}={\hat a}_{\bf p}$, with ${\bf p}\ne 0$, one instead finds the result 
\begin{equation}
n_{\bf p} = \langle \hat{a}^\dagger_{\bf p} \hat{a}_{\bf p}\rangle=  \frac{p^2/2m +gn}{2 \epsilon(p)}- \frac{1}{2} \; 
\label{npB}
\end{equation}
for the particle distribution function. Notice that $n_{\bf p}$ identically vanishes in the absence of interactions ($g=0$). It gives rise to the infrared divergent behavior \cite{Gavoret,PS} $n_{\bf p} \to mc/2p$ as $p\to 0$ and yields the result $\delta N_0/N =(8/3\sqrt\pi)(na^3)^{1/2}$ for the quantum depletion of the condensate. The  quantum depletion has  been recently  measured in a uniform 3D Bose Einstein condensed gas \cite{Hadz},  confirming the prediction of Bogoliubov theory.   

\textit{\bf Equation for the field operator.}  

As already mentioned in the introduction, time dependent Gross-Pitaevskii theory is well suited to study the dynamic response of the system to  space and time dependent external fields. In order to formulate the problem in the general context it is useful to derive the Gross-Pitaevskii theory starting from the Heisenberg equation  
\begin{equation}
i\hbar \frac{\partial}{\partial t}\hat{\Psi}({\bf r},t)= [\hat{\Psi}({\bf r},t),\hat{H}+\hat{H}_{pert}]
\label{GPQ}
\end{equation}
for the time evolution of the field operator, where  $\hat{H}_{pert}$ is the perturbative term (\ref{pert}).

The commutator involving the unperturbed Hamiltonian (\ref{H}) gives the result
\begin{equation}
[\hat{\Psi}({\bf r},t),\hat{H}]= \left[-\frac{\hbar^2 \nabla^2}{2m} +V_{ext} +g {\hat \Psi}^\dagger({\bf r}, t){\hat \Psi}({\bf r}, t) -\mu \right]\hat{\Psi}({\bf r},t) 
\label{GPQ1}
\end{equation}
where, for sake of generality, we have included an external trapping potential.  

In order to include the effect of the perturbation it is convenient to write ${\hat H}_{pert}$ in terms of the field operator. In the case of the coupling with the $\bf q$-component of the density operator one  writes  $\hat{F}=\hat{\rho}_{\bf q}=\int d{\bf r} {\hat \Psi}^\dagger({\bf r}){\hat \Psi}({\bf r})e^{-i{\bf q} \cdot {\bf r}}$  and the relevant  commutator takes the form
\begin{equation}
[\hat{\Psi}({\bf r},t),\hat{H}_{pert}]= -\lambda e^{\eta t}\left(e^{+i({\bf q} \cdot {\bf r}-\omega t)} +e^{-i({\bf q} \cdot {\bf r}-\omega t)} \right)\hat{\Psi}({\bf r},t) \; .
\label{GPQ2}
\end{equation} 
In the case of the coupling with the  ${\bf p}$-component of the field operator one chooses $\hat{F}={\hat \Psi}({\bf p})= (2\pi\hbar)^{-3/2}\int d{\bf r} e^{-i{\bf p} \cdot {\bf r}/\hbar} {\hat \Psi}({\bf r})$ and the relevant commutator instead becomes
\begin{equation}
[\hat{\Psi}({\bf r},t),\hat{H}_{pert}]= -\lambda e^{\eta t}(2 \pi \hbar)^{-3/2}e^{i({\bf p} \cdot {\bf r}/\hbar-\omega t)} \; .
\label{GPQ3}
\end{equation}

We are now ready to study the response function in the framework of Gross-Pitaevskii theory, where the field operator is replaced by a classical field. 

\textit{\bf Density response in Gross-Pitaevskii theory}  

By  replacing the field operator $\hat{\Psi}({\bf r})$ with the classical field $\Psi({\bf r})$ in  Eqs. (\ref{GPQ},\ref{GPQ1},\ref{GPQ2})  one obtains the time dependent Gross-Pitaevskii equation  
\begin{align}
&i\hbar \frac{\partial}{\partial t} \Psi({\bf r},t) =\left(-\frac{\hbar^2 \nabla^2}{2m} +V_{ext}+g |\Psi({\bf r}, t)|^2 -\mu \right) \Psi({\bf r},t) \nonumber \\ 
& -\lambda \sqrt{n({\bf r})}  e^{\eta t}\left(e^{i({\bf q} \cdot {\bf r}-\omega t)}+e^{-i({\bf q} \cdot {\bf r}-\omega t)}\right)
\label{GPrho}
\end{align}
in the presence of  the density  perturbation, where, in the last term  of the equation, we have taken the unperturbed value  $\Psi({\bf r},t)=\sqrt{n({\bf r})}$, consistent  with the rules of perturbation theory. 

In uniform conditions  the ansatz
\begin{equation} 
\Psi({\bf r},t)= \Psi_0 + e^{\eta t}\left(u e^{i({\bf q}\cdot {\bf r}-\omega t)} +v^*e^{-i({\bf q}\cdot {\bf r}-\omega t)}\right)
\label{ansatzrho}
\end{equation}
solves  the time dependent Gross-Pitaevskii equation  both in the absence and in the presence of the external density perturbation. In the above equation $\Psi_0$ is the  order parameter calculated at equilibrium.   In the absence of the external  perturbation one  finds the well known oscillating solutions with frequency 
\begin{equation}
\omega(q) = \sqrt{\frac{gn}{m} q^2+\hbar^2 (\frac {q^2}{2m})^2} .
\label{GPspectrum}
\end{equation}
This result is fully consistent with the dispersion relation (\ref{Bogspectrum}) predicted by Bogoliubov theory, after adopting the de Broglie quantization rules $\epsilon({\bf p})=\hbar \omega({\bf q})$ and ${\bf p}=\hbar {\bf q}$.  In the presence of the periodic density perturbation,   Eq. (\ref{GPrho}) can be also solved analytically, yielding  the following result for the amplitudes $u$ and $v$:
\begin{align}
&u= -\lambda \sqrt{n}\frac{\hbar \omega + p^2/2m}{(\hbar \omega+i\eta)^2 - \epsilon^2(p)} \nonumber
 \\
&v= -\lambda \sqrt{n}\frac{-\hbar \omega + p^2/2m}{(\hbar \omega+i\eta)^2 - \epsilon^2(p)} 
\label{uv}
\end{align}
where $\epsilon(p)$ is the Bogoliubov dispersion  law  (\ref{Bogspectrum})  and we have   used $\mu=gn$,  with $n=N/V$ the density of the system. Evaluating the variation  $\delta \rho^*_{\bf q}= \sqrt{n}\int d{\bf r} e^{-i{\bf q}\cdot {\bf r}}(\delta \Psi({\bf r},t)+\delta\Psi^*({\bf r},t))$ induced by the perturbation and using definition (\ref{chi}), we finally obtain the result 

\begin{align}
&\chi_{density}({\bf q},\omega)= -N\frac{p^2/m}{(\hbar\omega+i\eta)^2- \epsilon^2(p)} =\nonumber \\
&-N\left[\frac{1}{\hbar \omega +i\eta-\epsilon(p)}-\frac{1}{\hbar \omega +i\eta+\epsilon(p)}\right]\frac{p^2/m}{2\epsilon(p)}
\label{chirhoP}
\end{align}
for the density-density response function of the uniform gas (${\bf p}=\hbar {\bf q}$). By taking the imaginary part of the response function and using the fluctuation-dissipation theorem (\ref{FD}) one    immediately recovers the Bogoliubov result (\ref{rhorho}) for the density fluctuations. Result (\ref{chirhoP}) keeps the same form in the canonical and in the grand canonical formalism since the excitation operator $\hat{\rho}_{\bf q}$ commutes with $\hat N$. In the canonical case the ansatz for the order parameter satisfying the time-dependent Gross-Pitaevskii equation is simply obtained by multiplying Eq.(\ref{ansatzrho}) by the factor $exp(-i\mu t)$.

\textit{\bf Particle response in Gross-Pitaevskii theory}  

By  replacing the field operator $\hat{\Psi}({\bf r})$ with the classical field $\Psi({\bf r},t)$ in Eqs. (\ref{GPQ},\ref{GPQ1},\ref{GPQ3})   one instead  obtains the time dependent Gross-Pitaevskii equation 
\begin{align}
&i\hbar \frac{\partial}{\partial t} \Psi({\bf r},t) = \left(-\frac{\hbar^2 \nabla^2}{2m} + V_{ext}+g |\Psi({\bf r}, t)|^2-\mu \right) \Psi({\bf r},t) \nonumber \\
& -\lambda \frac{1}{(2\pi \hbar)^{3/2}}e^{i({\bf p} \cdot {\bf r}/\hbar-\omega t)}e^{\eta t} \; .
\label{GPp}
\end{align}
 accounting for the coupling with the ${\bf p}$-component ${\hat F}={\hat \Psi}({\bf p})$ of   field operator. 
We can still use the ansatz  (\ref{ansatzrho}) to solve the GP equation and in this case we obtain the following result for the amplitudes $u$ and $v$: 
\begin{align}
&u= \frac{\lambda}{(2\pi\hbar)^{3/2}}\frac{gn}{(\hbar \omega+i\eta)^2 - \epsilon^2(p)} \nonumber\\
&v= \frac{\lambda}{(2\pi\hbar)^{3/2}}\frac{\hbar \omega -p^2/2m-gn}{(\hbar \omega+i\eta)^2 - \epsilon^2(p)} \; .
\label {uvp}
\end{align}
The response function is then determined  by evaluating the fluctuations induced in the ${\bf p}$-component of the classical field
$\delta \Psi^*({\bf p},t)= (2\pi \hbar)^{-3/2}\int d{\bf r}e^{i{\bf p} \cdot {\bf r}/\hbar}\delta \Psi^*({\bf r},t)$. In uniform conditions  it is convenient to write the  ${\bf p}$-component ${\hat \psi}({\bf p})$ of the field operator in terms of the particle annihilation operator as 
\begin{equation}
\hat{\Psi}({\bf p}) = \frac{\sqrt{V}}{(2\pi \hbar)^{3/2}}\hat{a}_{\bf p}
\label{psia}
\end{equation}
so that the response function $\chi_{field}({\bf p},\omega)$ relative to   field operator $\hat{F}={\hat \psi}({\bf p})$ in momentum space
can be expressed in terms of the response function $\chi_{particle}({\bf p},\omega)$  relative to the particle operator ${\hat F}=\hat{a}_{\bf p}$ as
\begin{equation}
\chi_{field}({\bf p},\omega)= \frac{V}{(2\pi\hbar)^3}\chi_{particle}({\bf p},\omega) \; .
\label{chifield}
\end{equation}

Using results (\ref{uvp}) for $u$ and $v$ one finally finds the  following result  for the particle response function:
\begin{align}
&\chi_{particle}({\bf p},\omega) =  \frac{\hbar \omega -p^2/2m -gn}{(\hbar \omega+i\eta)^2 - \epsilon^2(p)} =\nonumber\\
& = -\frac{1}{2\epsilon(p)}\left(\frac{p^2/2m+gn-\epsilon(p)}{\hbar \omega   +i\eta -\epsilon(p)} - \frac{p^2/2m+gn+\epsilon(p)}{\hbar \omega + i\eta   +\epsilon(p)}\right) \; ,
\label{chip}
\end{align}
yielding the expression 
\begin{align}
A({\bf p},\omega) = &  \frac{1}{2\epsilon(p)} [(p^2/2m+gn-\epsilon(p))\delta(\hbar \omega -\epsilon(p)) \nonumber\\
&- (p^2/2m+gn+\epsilon(p))\delta(\hbar \omega +\epsilon(p))] 
\label{A}
\end{align}
for the spectral function, corresponding to the imaginary part of $\chi$. Result (\ref{chip}) shows that in the grand canonical formalism,  the particle response function shares the same poles of the density response function (\ref{chirhoP}). Equations (\ref{chip},\ref{A}) can be easily recast  in the canonical form, by simply replacing the frequency $\omega$ with $\omega+\mu/\hbar$. This reflects the fact that the operator $\hat{a}^\dagger_{\bf p}$ ($\hat{a}_{\bf p}$)  add (remove) a particle, in addition to creating or annihilating  an elementary mode in the system. In the canonical formalism the solution for the order parameter would actually take the form
\begin{equation} 
\Psi({\bf r},t)= e^{-i\mu t}\Psi_0 + e^{\eta t}\left(u e^{-2i\mu t}e^{i({\bf q}\cdot {\bf r}-\omega t)} +v^*e^{-i({\bf q}\cdot {\bf r}-\omega t)}\right) \;. 
\label{ansatzrhoP}
\end{equation}

For  large values of  $\omega$ the response function approaches 
the value $1/(\hbar \omega)$ in agreement with the general result  
\begin{equation}
\chi_F(\omega \to \infty) =\frac{1}{\hbar \omega}\langle [{\hat F},{\hat F}^\dagger]\rangle
\end{equation}
holding for the dynamic polarizability in the large $\omega$ limit  \cite{book}, involving the commutator between ${\hat F}$ and ${\hat F}^\dagger$ (see Eq.(\ref{kubo2}).
 
Using the fluctuation dissipation theorem (\ref{FD+})    one  exactly recovers  the result (\ref{npB}) predicted by Bogoliubov theory for the particle distribution function, characterized by the infrared divergence $n_{\bf p} \to mc/p$ at small $p$ and accounting for the quantum depletion of the condensate. 

Analogously, one can also derive the expression for the mixed particle-density response function, 
providing the fluctuations induced in the average of the particle operator $\hat{a}_{\bf p}^\dagger$ by the presence of an external perturbation coupled to the density operator $\hat{\rho}_{\bf q}$ with ${\bf q}={\bf p}/\hbar$. Such a perturbation modifies the wave function of the condensate according to  Eqs. (\ref{ansatzrho},\ref{uv}). One finally finds 
\begin{align}
&\chi_{particle-density}({\bf p},\omega) =  \nonumber \\
&-\frac{1}{\hbar} \sum_{n}\left[\frac{\langle 0|{\hat a}^\dagger_{\bf p}|n\rangle\langle n |{\hat \rho}_{\bf q}|0\rangle}{\omega-\omega_{n}+i\eta}-\frac{\langle 0|{\hat \rho}_{\bf q}|n\rangle\langle n |{\hat a}^\dagger_{\bf p}|0\rangle}{\omega+\omega_{m}+i\eta}\right] = \nonumber \\ 
& \sqrt{N} \frac{\hbar \omega - p^2/2m}{(\hbar \omega+i\eta)^2 - \epsilon^2(p)}  .
\label{pd}
\end{align}
Result (\ref{pd}) is  consistent with the large $\omega$ result $\chi_{particle-density}\to -\langle [{\hat a}^\dagger_{\bf p},{\hat \rho}_{\bf q}]\rangle/(\hbar \omega)$,  derivable from sum rule arguments \cite{sandrocrossed}. In the canonical formalism the physical meaning of Eq. (\ref{pd}) would correspond to replacing the operator ${\hat a}^\dagger_{\bf p}$ with the number conserving operator ${\hat a}^\dagger_{\bf p} \hat{a}_0/\sqrt{N_0}$.

In the static limit the result $\chi_{particle-density}({\bf p},\omega=0)= \sqrt{N} (p^2/2m)/\epsilon^2(p)$ can be used to investigate the effect of a static periodic  perturbation of the form $H_{pert} = -\lambda ({ \hat{\rho}}_{\bf q}+\hat{\rho}_{-\bf q})= -2 \lambda \sum_j cos(qz_j)$ on the momentum  distribution of the system, which turns out to be characterized by   the occurrence  of  the macroscopic occupation $N_{\bf p}= N [\lambda (p^2/2m)/\epsilon^2(p)]^2$ of the single particle state with momentum ${\bf p}$ (and analogously for   $-\bf p$). The effect  should be observable  experimentally also for relatively small values of the coupling $\lambda$ in systems exhibiting a pronounced roton minimum as happens, under proper conditions,  in the case of long-range dipolar interactions \cite{santosrotons, santos,ferlaino}. The coupling between density and particle excitations accounted for by Eq. (\ref{pd}) reflects a peculiar property of a Bose-Einstein condensate and disappears in the absence of coherence, as proven experimentally for large intensities of the external density coupling when the system enters the insulator phase \cite{greiner}. 

\textit{\bf Response function in non uniform systems} 

The above results can be straightforwardly generalized to the case of a non uniform trapped Bose-Einstein condensed gas,  where the Hamiltonian contains an external static potential $V_{ext}$. In this case the density response function  takes the form :
\begin{equation}
\chi_{density}({\bf q},\omega)= -\sum_n\frac{|\int d{\bf r} (u_n({\bf r})+v_n({\bf r}))\Psi_0({\bf r})e^{-i{\bf q} \cdot {\bf r}}|^2}{(\hbar \omega+i\eta)^2- \epsilon^2_n} \; ,
\label{chirhoGP}
\end{equation}
while the response to the  field operator $F=\hat{\Psi}({\bf p})$ in momentum space reads
\begin{align}
&\chi_{field}({\bf p})= -\frac{1}{(2\pi \hbar)^3}\sum_n [\frac{|\int d{\bf r} v_n({\bf r})e^{-i{\bf p} \cdot {\bf r}/\hbar}|^2}{(\hbar \omega+ i\eta)- \epsilon_n} \nonumber \\
&- \frac{|\int d{\bf r} u_n({\bf r})e^{-i{\bf p} \cdot {\bf r}/\hbar}|^2}{(\hbar \omega + i\eta)- \epsilon_n} ] \; .
\label{chirhoGPfield}
\end{align}
In both Eqs. (\ref{chirhoGP}) and (\ref{chirhoGPfield})   $u_n$, $v_n$ and $\epsilon_n$ are provided by the solutions of the coupled Gross-Pitaevskii equations
 \begin{align}
&\epsilon_n u_n= \left(H_0 -\mu +2gn({\bf r})\right)u_n({\bf r}) +gn({\bf r})v_n({\bf r}) \nonumber \\
&-\epsilon_n v_n= \left(H_0 -\mu +2gn({\bf r})\right)v_n({\bf r}) +gn({\bf r})u_n({\bf r}) 
\end{align}
which are the analogs of the Bogoliubov equations of uniform matter and the sum over $n$   includes all the excitations of the system. Here $H_0= -(\hbar^2/2m) \nabla^2 + V_{ext}$ is the single-particle Hamiltonian. The   amplitudes $u_n$ and $v_n$ satisfy the ortho-normalization condition $\int d{\bf r} (u_n^*u_m -v_n^*v_m) = \delta_{nm}$ and in a uniform gas take the form  $u_n\equiv u_{\bf k}({\bf r})= u_{\bf k}e^{i{\bf k}\cdot {\bf r}}$, and analogously for $v_n$.

Equation (\ref{chirhoGPfield}), together with result (\ref{FD+}), allows for the calculation of the momentum distribution 
\begin{align} 
n({\bf p}) &= \langle {\hat \Psi}^\dagger({\bf p}){\hat \Psi}({\bf p})\rangle = \nonumber \\
&= |\Psi_0({\bf p)}|^2 + \frac{1}{(2\pi \hbar)^3} \sum_n |\int d{\bf r} v_n({\bf r})e^{-i{\bf p} \cdot {\bf r}/\hbar}|^2  \; .
\label{npfinal}
\end{align}
In addition to the mean field contribution $|\Psi_0({\bf p})|^2$, fixed by the Fourier transform $\Psi_0({\bf p})=(2\pi\hbar)^{-3/2}\int d{\bf r}e^{i{\bf p} \cdot {\bf r}/\hbar}\Psi_0({\bf r})$ of the order parameter at equilibrium and providing the leading contribution for $p < \hbar/R$ with  $R$  the typical size of the condensate, Eq. (\ref{npfinal}) accounts for the  quantum fluctuations caused by the elementary excitations of the system and provides  the leading contribution at larger values of $\bf p$. 
The experimental determination of $n({\bf p})$ for large values of ${\bf p}$ has been the object of a recent time-of-flight  investigation \cite{clement}. The presence of interactions during the expansion does not however allow, in this experiment,  for a safe identification of the in-situ momentum distribution \cite{qu}.

Another instructive example concerns the calculation of the fluctuations of the field operator ${\hat F}=\hat{\Psi}({\bf r})$ in coordinate space. In this case one finds the result 
\begin{align} 
n({\bf r}) &= \langle {\hat \Psi}^\dagger({\bf r}){\hat \Psi}({\bf r})\rangle = \nonumber \\
&= |\Psi_0({\bf r)}|^2 + \sum_n | v_n({\bf r})|^2  \; ,
\label{nrfinal}
\end{align}
which provides a natural decomposition of the density into the Gross-Pitaevskii value $|\Psi_0({\bf r)}|^2$ and the contribution arising from the fluctuations of the condensate. In uniform configurations  the values of $v_{\bf p}$ are fixed by Eq.(\ref{uv}) and the  decomposition corresponds to writing $N=N_0+\delta N_0$ with $\delta N_0= \sum_{\bf p} | v_p|^2 =N (8/3\sqrt\pi)(na^3)^{1/2}$. In non uniform configurations  the use of Eq.(\ref{nrfinal}) requires more careful considerations. In fact while the fluctuations of the field operator are   proportional to the perturbation  parameter that scales as $a^{3/2}$, the order parameter $\Psi_0$, calculated in GP theory, ignores corrections of the same order arising from the renormalization of the coupling constant, as predicted by the theory of Lee-Huang-Yang \cite{LHY}. By evaluating the order parameter $\Psi_0$  using the Gross-Pitaevskii theory in the Thomas-Fermi (LDA) approximation,  one can  in fact easily show that the prediction of  (\ref{nrfinal}) differs from the total density derivable by including the  LHY correction in the equation of state \cite{timmermans} (see also \cite{book}, Sect. 11.5).   It is worth noticing that both the LHY   and the fluctuation correction affect the density profile in the same physical region where $r <R_{TF}$ and  the density   significantly differs from zero. This differs from the case of the momentum distribution where, as already pointed out, the fluctuations of the condensate modify the momentum distribution in the region $p> \hbar/R_{TF}$ where the value of $\Psi_0({\bf p})$ is negligible. 

\text{\bf Chemical potential and beyond mean field effects} 

The Gross-Pitaevskii equation for the order parameter (see Eqs. (\ref{GPrho}) and (\ref{GPp})) has been derived replacing the field operator $\hat{\Psi}$ with the classical field $\Psi$ in the  equation for the field operator (\ref{GPQ}). This procedure, when applied to the average of Eq. (\ref{GPQ}), ignores the presence of fluctuations in the quantity $\langle\hat{\Psi}^\dagger({\bf r},t)\hat{\Psi}({\bf r},t)\hat{\Psi}({\bf r},t)\rangle$,  which can be conveniently witten in the form:
\begin{align}
\langle\hat{\Psi}^\dagger({\bf r},t)\hat{\Psi}({\bf r},t)\hat{\Psi}({\bf r},t)\rangle &= 
\langle\hat{n}({\bf r},t)\rangle \langle \hat{\Psi}({\bf r},t)\rangle  \nonumber \\
& + \langle\delta \hat{n}({\bf r},t)\delta \hat{\Psi}({\bf r},t)\rangle 
\label{psi3}
\end{align} 
with $\hat{n} = \hat{\Psi}^\dagger \hat{\Psi}$,  $\delta\hat{n}= \hat{n}-\langle \hat{n}\rangle$ and $\delta \hat{\Psi}= \hat{\Psi}-\langle\hat{\Psi}\rangle$.
The first term in the RHS of the above equation  coincides with the quantity $n({\bf r},t) \Psi({\bf r},t)$ and, neglecting quantum depletion effects in the density, i.e. setting $n({\bf r},t) =\Psi^*({\bf r},t) \Psi({\bf r},t)$, provides the usual interaction term entering the Gross-Pitaevskii equation. The second term is instead associated with the density-particle fluctuations discussed in the previous part of the paper (see Eq. (\ref{pd})) and is ignored in the derivation of the Gross-Pitaevskii equation.   By explicitly accounting for these   fluctuations one can  improve the   equation for the order parameter  in a perturbative way accounting for beyond mean field effects \cite{castin}.  

A first important result is obtained by identifying the stationary solution in uniform matter and  in the absence of external perturbations. By writing 
$\delta \hat{n}({\bf r})= (1/V) \sum_{\bf q}e^{i{\bf q}\cdot {\bf r}}\hat{\rho}_{\bf q}$ and $\delta \hat{\Psi}({\bf r})= (1/\sqrt{V}) \sum_{\bf p}e^{i{\bf p}\cdot {\bf r}/\hbar} \hat{a}_{\bf p}$,  and noticing that in uniform matter only the terms ${\bf p}=-\hbar {\bf q}$ give non vanishing contributions,  
 the equation for the order parameter $\Psi_0$ takes the  form:
\begin{align}
\mu \Psi_0 &= gn\left(1 +\frac{g}{V}\sum_{{\bf p}\ne 0}\frac{m}{p^2}\right)\Psi_0 \nonumber \\
& + g\frac{1}{V\sqrt{V}} \sum_{{\bf p}\ne 0}\langle \hat{\rho}_{-{\bf p}/\hbar}\hat{a}_{{\bf p}}\rangle \; ,
\label{muS}
\end{align}
 where, consistently with the beyond mean field procedure, we have taken into account the renormalization $g \to g(1 +g/V\sum_{{\bf p}\ne 0}m/p^2)$ of the coupling constant, avoiding the occurrence of ultraviolet divergences.  Using the identity $\langle \hat{\rho}_{-{\bf p}/\hbar}\hat{a}_{{\bf p}}\rangle= \langle \hat{a}_{{\bf p}}^\dagger\hat{\rho}_{{\bf p}/\hbar}\rangle^*=\langle \hat{a}_{{\bf p}}^\dagger\hat{\rho}_{{\bf p}/\hbar}\rangle$ and Eq. (\ref{pd}) for the particle-density response function, one easily finds the result $\langle \hat{\rho}_{-{\bf p}/\hbar}\hat{a}_{{\bf p}}\rangle= \sqrt{N}/(2\epsilon(p)) (p^2/2m-\epsilon(p))$, where $\epsilon(p))$ is the Bogoliubov expression (\ref{Bogspectrum}) for the energy of the elementary excitations carrying momentum $p$. By further replacing the  quantity $\sqrt{N/V}$ with the order parameter $\Psi_0$ in the last term of Eq. (\ref{muS}), one finally obtains  the relevant  result 
\begin{align}
\mu &= gn + \frac{g}{(2\pi\hbar)^3}\int d{\bf p} \left[ \frac{p^2/2m-\epsilon(p)}{2 \epsilon(p)} + \frac{gn m}{p^2}\right] = \nonumber \\
& = gn\left[ 1+ \frac{32}{3\sqrt\pi}(na^3)^{1/2}\right] 
\label{muSS}
\end{align}
for the chemical potential, which  includes the first   correction to the mean field value $\mu=gn$.
Result (\ref{muSS}) coincides with the value derivable from the Lee-Huang-Yang expression 
\begin{equation}
E_{0}/V = \frac{1}{2}gn^2 +\frac{1}{2(2\pi\hbar)^3} \int d{\bf p}\left[\epsilon(p) - gn -\frac{p^2}{2m} +(gn)^2\frac{m}{p^2}\right]
\label{ELHY}
\end{equation}
for the ground state energy 
as can be explicitly checked  using the thermodynamic relation $\mu= \partial E_{0}/\partial N$. The Lee-Huang-Yang energy   is usually calculated  through a proper diagonalization of the Bogoliubov Hamiltonian (see derivation of Eq. (\ref{ELHY+})),  as well as taking into account the renormalization of the interaction coupling constant, so that the present derivation provides a further insightful  link between the Bogoliubov formalism and the one based on  the equation for the order parameter.  

\textit{\bf Conclusions}.  
In conclusion we have shown that the use of the $T=0$ limit of the fluctuation dissipation theorem allows for the calculation of the quantum fluctuations of both the density and particle operators of a Bose-Einstein condensed gas, employing the time dependent Gross-Pitaevskii equation  for the  wave function of the condensate, a classical field describing the order parameter of the system. This approach enlightens   the deep equivalence between the Bogoliubov and Gross-Pitaevskii approaches, despite their different theoretical formulation. The suitability of the GP approach to describe  non uniform configurations might offer novel possibilities for investigating the nature of the fluctuations in the presence of quantum defects, like solitons and quantized vortices.  We have also shown that the calculation  of the density-particle fluctuations  permits to generalize  the equation for the order parameter, allowing  for the determination of the chemical potential beyond the mean field picture, in agreement with the predictions of  Lee-Huang-Yang theory.

\textit{\bf Acknowledgments}.
It is a great pleasure to thank long-standing scientific collaborations and stimulating discussions  with Lev Pitaevskii, which started 30 years ago, after my first visit to the Kapitza Institute for Physical Problems in Moscow and are still now   continuing successfully in Trento.


\begin{thebibliography}{52}
\bibitem{FD} At finite temperature the function $sign(\omega)$ should be replaced by $\coth(\beta \hbar \omega/2)$.
\end{thebibliography}
\end{document}